\documentclass[amsmath,amssymb,amsfonts,aps,pre,preprint,superscriptaddress,bibnotes,showpacs,showkeys,longbibliography]{revtex4-1}

\usepackage{graphicx}
\usepackage{subfigure}
\usepackage{amsmath}
\usepackage{amsfonts}
\usepackage{amssymb}
\usepackage{natbib}
\usepackage{relsize}
\usepackage{sidecap}
\usepackage{braket}
\usepackage{footmisc}
\usepackage{footnote}
\usepackage{color}
\begin{document}

\title{Density of condensate of a dilute Bose gas in improved Hartree-Fock approximation}

\author{Nguyen Van Thu}
\affiliation{Institute for Research and Development, Duy Tan University, Da Nang, Vietnam}
\affiliation{Department of Physics, Hanoi Pedagogical University 2, Hanoi, Vietnam}

\begin{abstract}
In a manner of Cornwal-Jackiw-Tomboulis effective action approach, the density of condensate of a dilute Bose gas confined between two planar walls is investigated within the framework of improved Hartree-Fock. Thereby, the quantum fluctuations are taken into account with presence of high-order terms in the momentum integrals. Our results show that the quantum fluctuations significantly belittle the density of condensate compared with square of the expectation value of the field operator. The comparison with relating results of Gross-Pitaevskii theory is made. 
\end{abstract}

\maketitle

\section{Introduction\label{sec1}}

The density of condensate of a Bose gas (BEC) plays an important role in studies on BEC in both theory and experiment. Once the density of condensate is found, all of static and dynamic properties of the BEC can be easily investigated.

In Gorss-Pitaevskii (GP) theory, the BEC is assumed at zero temperature therefore all of particles are condensed. The ground state is described by the wave function, which is the solution of a nonlinear differential equation called the GP equation. The density of condensate is defined as square of the wave function and called the profile \cite{Pitaevskii,Pethick}. However, it is reported that we have never achieve the absolute zero temperature and even so, a number of particles will be in the excited state instead of the ground state \cite{Annett}. Based on the GP theory, the density of condensate in the BEC have been investigated by many authors, for example, see Refs. \cite{Aochui,Brankov,Mazets} and so on. In a simple approximation called double-parabola approximation \cite{Joseph} and triple-parabola approximation \cite{Joseph1}, this aspect was also studied.

Taking into account the quantum fluctuations, quantum field theory is proposed in several levels of the accuracy. In the one-loop approximation, the density of condensate in the ideal and weakly interacting Bose gas was investigated \cite{Andersen}. The Cornwal-Jackiw-Tomboulis (CJT) effective action approach \cite{CJT} can be applied to consider the contribution of two-loop diagrams. The authors of Ref. \cite{Phat} employed this method to consider the scenarios of phase transition in a binary mixture of Bose gases (BECs).

The finite-size effect in the BEC(s) produces many amazing changes in properties of the BEC(s). Within the framework of the GP theory, the density of condensate of the BEC has been researched \cite{Thu382,Biswas3} and \cite{Thuphatsong,Takahashi} for BECs. One of the most interesting consequence of the finite-size effect is the Casimir effect, which has been studied widely in many scopes of physics. In the BEC field, this effect was investigated in the one-loop approximation \cite{Thu382,Biswas3,Schiefele}, in which the expectation value of the field operator is assumed independence of the coordinate. The results showed that the expectation value of the field operator is not only independent of the coordinate but also of the size of system, which is usually the distance between two planar walls filled by the Bose gas. By the same way, this effect was also studied in BECs \cite{Thu1}. In higher approximation, called improved Hartree-Fock approximation, tthe expectation value of the field operator was considered in BECs with lower-order terms in the momentum integrals \cite{ThuIJMPB}. The contribution of the higher-order terms is taken into account to investigated the density of condensate of the BEC \cite{ThuPhysa}. The resulting the expectation value of the field operator depends on both the coordinate and the distance between two planar walls. In this paper, the density of condensate of the dilute Bose gas confined between two parallel plates is considered in the improved Hartree-Fock approximation with the present of the high-order terms in the momentum integrals.

This paper is organized as follows. In Section \ref{sec2}, the density of condensate is derived in improved Hartree-Fock approximation. Section \ref{sec3} devote the density of condensate of a weakly interacting Bose gas in HIHF approximation. Conclusions are given in Section \ref{sec4}.

\section{Density of condensate in the improved Hartree-Fock approximation\label{sec2}}

We start by considering a dilute Bose gas, which is described by the Lagrangian \cite{Pethick},
\begin{eqnarray}
{\cal L}=\psi^*\left(-i\hbar\frac{\partial}{\partial t}-\frac{\hbar^2}{2m}\nabla^2\right)\psi(\vec{r},t)-\mu\left|\psi(\vec{r},t)\right|^2+\frac{g}{2}\left|\psi(\vec{r},t)\right|^4,\label{1}
\end{eqnarray}
in which $\psi(\vec{r},t)$ is the field operator and its expectation value plays the role of the order parameter; the chemical potential and temperature are denoted by $\mu$ and $T$, respectively; $\hbar$ is the reduced Plack's constant. The strength of interaction among atoms is determined via the $s$-wave scattering length $a_s$ in the form
\begin{eqnarray}
g=4\pi\hbar^2a_s/m>0,\label{g}
\end{eqnarray}
for the repulsive interaction with $m$ being the atomic mass.

Let $\psi_0$ be the expectation value of the field operator in the tree approximation, in broken phase one has
\begin{eqnarray}
\psi_0^2=\frac{\mu}{g}.\label{dentree}
\end{eqnarray}
In momentum space, the inversion propagator is read off
\begin{eqnarray}
D_0^{-1}(k)&=&\left(
              \begin{array}{cc}
                \frac{\hbar^2k^2}{2m}+2g\psi_0^2 & -\omega_n \\
                \omega_n &  \frac{\hbar^2k^2}{2m}\\
              \end{array}
            \right),\label{protree}
\end{eqnarray}
in which $\vec{k}$ is the wave vector and $\omega_n=2\pi n/\beta,~n=0,\pm1,\pm2,...$ is the Matsubara frequency for boson. At temperature $T$ one has $\beta=1/k_BT$ with $k_B$ being the Boltzmann constant. The vanishing of the determinant \cite{Floerchinger} $\det D_0^{-1}(k)=0$ allows us to obtain the dispersion relation
\begin{eqnarray}
E(k)=\sqrt{\frac{\hbar^2k^2}{2m}\left(\frac{\hbar^2k^2}{2m}+2g\psi_0^2\right)}.\label{dispertree}
\end{eqnarray}
It is obvious that there is a Goldstone boson associated with $U(1)$ breaking.

Now the quantum fluctuations are taken into account by expanding the field operator in terms of its expectation value and two real fields $\psi_1,\psi_2$ associated with the quantum fluctuations \cite{Andersen},
\begin{eqnarray}
\psi\rightarrow \psi_0+\frac{1}{\sqrt{2}}(\psi_1+i\psi_2).\label{shift}
\end{eqnarray}
Putting (\ref{shift}) into Lagrangian (\ref{1}) one has the interacting Lagrangian in double-bubble approximation \cite{ThuIJMPB,ThuPhysa},
\begin{eqnarray}
{\cal L}_{int}=\frac{g}{2}\psi_0\psi_1(\psi_1^2+\psi_2^2)+\frac{g}{8}(\psi_1^2+\psi_2^2)^2.\label{Lint}
\end{eqnarray}
The Cornwal-Jackiw-Tomboulis (CJT) effective potential can be read off from (\ref{Lint})
\begin{eqnarray}
V_\beta=&&-\mu\psi_0^2+\frac{g}{2}\psi_0^4+\frac{1}{2}\int_\beta \mbox{tr}\left[\ln D^{-1}(k)+D_0^{-1}(k)D(k)-{1\!\!1}\right]\nonumber\\
&&+\frac{3g}{8}(P_{11}^2+P_{22}^2)+\frac{g}{4}P_{11}P_{22},\label{VHF}
\end{eqnarray}
with ${1\!\!1}$ being a unit matrix, $D(k)$ is the propagator in the double-bubble approximation and notations
\begin{eqnarray*}
&&\int_\beta f(k)=\frac{1}{\beta}\sum_{n=-\infty}^{+\infty}\int\frac{d^3\vec{k}}{(2\pi)^3}f(\omega_n,\vec{k}),\nonumber\\
&&P_{11}=\int_\beta D_{11}(k),~P_{22}=\int_\beta D_{22}(k),\label{Paa}
\end{eqnarray*}
are introduced. In our previous papers \cite{ThuIJMPB,ThuPhysa}, the CJT effective potential (\ref{VHF}) was proved not satisfying the Goldstone theorem.  In order to restore it, the method proposed by Ivanov et. al. \cite{Ivanov} is invoked therefore instead of (\ref{VHF}) one has a new CJT effective potential \cite{ThuPhysa},
\begin{eqnarray}
\widetilde{V}_\beta=&&-\mu\psi_0^2+\frac{g}{2}\psi_0^4+\frac{1}{2}\int_\beta \mbox{tr}\left[\ln D^{-1}(k)+D_0^{-1}(k)D(k)-{1\!\!1}\right]\nonumber\\
&&+\frac{g}{8}(P_{11}^2+P_{22}^2)+\frac{3g}{8}P_{11}P_{22}.\label{VIHF}
\end{eqnarray}
It is not difficult to verify that the CJT effective potential (\ref{VIHF}) reproduces the Goldstone boson with a new dispersion relation \cite{ThuPhysa},
\begin{eqnarray}
E(k)=\sqrt{\frac{\hbar^2k^2}{2m}\left(\frac{\hbar^2k^2}{2m}+M^2\right)}.\label{disperIHF}
\end{eqnarray}
with $M$ being the effective mass and the approximation associated with the effective potential (\ref{VIHF}) is called the IHF approximation. The inversion propagator is now
\begin{eqnarray}
D^{-1}(k)=\left(
              \begin{array}{lr}
                \frac{\hbar^2k^2}{2m}+M^2 & -\omega_n \\
                \omega_n & \frac{\hbar^2k^2}{2m} \\
              \end{array}
            \right).\label{proIHF}
\end{eqnarray}
Eqs. (\ref{Paa}) and (\ref{proIHF}) give us the momentum integrals
\begin{eqnarray}
&&P_{11}=\frac{1}{2}\int\frac{d^3\vec{k}}{(2\pi)^3}\sqrt{\frac{\hbar^2k^2/2m}{\hbar^2k^2/2m+M^2}},\nonumber\\
&&P_{22}=\frac{1}{2}\int\frac{d^3\vec{k}}{(2\pi)^3}\sqrt{\frac{\hbar^2k^2/2m+M^2}{\hbar^2k^2/2m}}.\label{tichphan}
\end{eqnarray}
From (\ref{VIHF}) it is easy to derive the gap equation
\begin{eqnarray}
-\mu+g\psi_0^2+\frac{3g}{2}P_{11}+\frac{g}{2}P_{22}=0,\label{gap}
\end{eqnarray}
and Schwinger-Dyson (SD) equation
\begin{eqnarray}
M^2=-\mu+3g\psi_0^2+\frac{g}{2}P_{11}+\frac{3g}{2}P_{22}.\label{SD}
\end{eqnarray}
 The pressure is defined as
\begin{eqnarray}
{\cal P}=-\widetilde{V}_\beta\bigg|_{\mbox{at minimum}},\label{press}
\end{eqnarray}
and thus the density of condensate in the IHF approximation is determined by
\begin{eqnarray}
\rho=-\frac{\partial {\cal P}}{\partial \mu}.\label{rho}
\end{eqnarray}
Combining Eqs. (\ref{VIHF}) and (\ref{gap})-(\ref{rho}) one obtains the density of condensate in the IHF approximation
\begin{eqnarray}
\rho_{{\text{IHF}}}=\psi_0^2+\frac{1}{2}(P_{11}+P_{22}).\label{roIHF}
\end{eqnarray}

\section{Density of condensate of a weakly interacting Bose gas in HIHF approximation\label{sec3}}

Consider a weakly interacting Bose gas confined between two hard walls, these walls are perpendicular to $0z$ axis and separating at distance $\ell$.  Along
$0x, 0y$ directions, the system under consideration is translational. Because of the compactification in $z$-direction, the wave vector is quantized
\begin{eqnarray}
k^2\rightarrow k_\perp^2+k_j^2,\label{k}
\end{eqnarray}
in which the wave vector component $k_\perp$ is perpendicular to $0z$-axis and $k_j$ is parallel with $0z$-axis. Impose the periodic boundary condition at the hard walls \cite{ThuPhysa}, the $k_j$ component of the wave vector has the form
\begin{eqnarray}
k_j=\frac{2\pi j}{\ell},~j\in{\mathbb{Z}}.\label{kj}
\end{eqnarray}
In order to simplify notations, the distance between two plates is scaled by the healing length $\xi=\hbar/\sqrt{2mgn_0}$ with $n_0$ being the bulk density thus the dimensionless distance is $L=\ell/\xi$. The dimensionless wave vector is $\kappa=k\xi$ and Eq. (\ref{k}) becomes
\begin{eqnarray}
\kappa^2\rightarrow\kappa_\perp^2+\kappa_j^2,\label{kappaj}
\end{eqnarray}
with the dimensionless form of (\ref{kj}) being $\kappa_j=j/\overline{L},~\overline{L}=L/(2\pi)$. Similarly, the momentum integrals (\ref{tichphan}) can be transformed into the dimensionless form
\begin{eqnarray}
&&P_{11}=\frac{1}{2\xi^3}\int\frac{d^3\vec{\kappa}}{(2\pi)^3}\frac{\kappa}{\sqrt{\kappa^2+{\cal M}^2}},~P_{22}=\frac{1}{2\xi^3}\int\frac{d^3\vec{\kappa}}{(2\pi)^3}\frac{\sqrt{\kappa^2+{\cal M}^2}}{\kappa}.\label{tichphan1a}
\end{eqnarray}
Because of the quantization of the wave vector (\ref{kappaj}), the momentum integrals (\ref{tichphan1a}) become
\begin{eqnarray}
P_{11}&=&\frac{1}{2\xi^3}\sum_{j=-\infty}^{+\infty}\int_0^\Lambda \frac{d^2\kappa_\perp}{(2\pi)^2}\sqrt{\frac{\kappa_\perp^2+\kappa_j^2}{\kappa_\perp^2+\kappa_j^2+{\cal M}^2}},\nonumber\\
P_{22}&=&\frac{1}{2\xi^3}\sum_{j=-\infty}^{+\infty}\int_0^\Lambda \frac{d^2\kappa_\perp}{(2\pi)^2}\sqrt{\frac{\kappa_\perp^2+\kappa_j^2+{\cal M}^2}{\kappa_\perp^2+\kappa_j^2}},\label{k2}
\end{eqnarray}
where the dimensionless effective mass is defined as ${\cal M}=M/\sqrt{gn_0}$. Because of the ultraviolet divergence of the integrating over the perpendicular component of the wave vector, a cut-off $\Lambda$ is introduced in Eqs. (\ref{k2}). By this way these integration can be performed. The remaining is the sum over the parallel component $\kappa_j$ of the wave vector. To deal with this sum, the Euler-Maclaurin formula \cite{Arfken} is employed
\begin{eqnarray}
\sum_{n=0}^\infty \theta_nF(n)-\int_0^\infty F(n)dn=-\frac{1}{12}F'(0)+\frac{1}{720}F'''(0)-\frac{1}{30240}F^{(5)}(0)+\cdots,\label{EM}
\end{eqnarray}
with
\begin{eqnarray*}
\theta_n=\left\{
           \begin{array}{ll}
             1/2, & \hbox{if $n=0$;} \\
             1, & \hbox{if $n>0$.}
           \end{array}
         \right.
\end{eqnarray*}
After taking the sum, let the cut-off $\Lambda$ tends to infinity one has \cite{ThuPhysa},
\begin{eqnarray}
P_{11}=-\frac{mgn_0\pi^2\xi^2}{90\hbar^2{\cal M}\ell^3}, ~P_{22}=\frac{mgn_0{\cal M}}{12\hbar^2\ell}-\frac{mgn_0\pi^2\xi^2}{90\hbar^2{\cal M}\ell^3}.\label{k4}
\end{eqnarray}
In comparison with those in Refs. \cite{ThuPhysa}, the higher-order terms are taken into account in this paper, this approximation will be called the higher-order improved Hartree-Fock (HIHF) approximation. In case of absence these terms, the approximation calls lower-order improved Hartree-Fock (LIHF) approximation.

Next we calculate the density of condensate in the HIHF approximation and evaluate the contribution of the higher-order terms in the density of condensate. According to Eq. (\ref{roIHF}), one first finds the order parameter in the HIHF approximation. To do that, the gap and SD equations (\ref{gap}) and (\ref{SD}) should be converted into the dimensionless form, in which the effect from compactification of the wave vector has to be noticed. Plugging (\ref{k4}) into (\ref{gap}) and (\ref{SD}) one arrives at
\begin{eqnarray}
-1+\phi^2+\frac{mg{\cal M}}{24\xi\hbar^2L}-\frac{mg\pi^2}{45\xi\hbar^2{\cal M}L^3}&=&0,\label{gap1}
\end{eqnarray}
for the gap equation and
\begin{eqnarray}
-1+3\phi^2+\frac{mg{\cal M}}{8\xi\hbar^2L}-\frac{mg\pi^2}{45\xi\hbar^2{\cal M}L^3}&=&{\cal M}^2,\label{SD1}
\end{eqnarray}
for the SD equation. Note that here we use the dimensionless distance $L=\ell/\xi$ and the reduced order parameter $\phi=\psi_0/\sqrt{n_0}$. To simplify the solution of the gap and SD equations, the gas parameter $n_s=n_0a_s^3$ is introduced. It is well-known that a Bose gas is dilute if the gas parameter satisfies the condition $n_s\ll1$. Combining Eqs. (\ref{gap1}), (\ref{SD1}) and (\ref{g}), the gap and SD equations can be written in term of the gas parameter
\begin{eqnarray}
-1+\phi^2+\frac{\sqrt{2}\pi^{3/2}n_s^{1/2}{\cal M}}{3L}-\frac{8\sqrt{2}\pi^{7/2}n_s^{1/2}}{45L^3{\cal M}}&=&0,\nonumber\\
-1+3\phi^2+\frac{\sqrt{2}\pi^{3/2}n_s^{1/2}{\cal M}}{L}-\frac{8\sqrt{2}\pi^{7/2}n_s^{1/2}}{45L^3{\cal M}}&=&{\cal M}^2.\label{gapSD}
\end{eqnarray}
Introducing a parameter $\alpha$  \cite{ThuPhysa},
\begin{eqnarray}
\cos\alpha=\frac{4\pi^{7/2}n_s^{1/2}}{\sqrt{75}L^3},\label{alpha}
\end{eqnarray}
the solution for Eqs. (\ref{gapSD}) can be written in form
\begin{eqnarray}
{\cal M}&=&\sqrt{\frac{2}{3}}\cos\frac{\alpha}{3},\nonumber\\
\phi^2&=&1+\frac{8 \pi ^{7/2}n_s^{1/2} \sec \frac{\alpha }{3}}{15 \sqrt{3} L^3}-\frac{2 \pi ^{3/2} n_s^{1/2} \cos \frac{\alpha }{3}}{3 \sqrt{3} L}.\label{nghiem}
\end{eqnarray}
Now we put Eqs. (\ref{nghiem}) and (\ref{k4}) into the definition of the density of condensate one has
\begin{eqnarray}
\varrho_{\text{HIHF}}=1+\frac{4 \pi ^{7/2} n_s^{1/2} \sec \left(\frac{\alpha }{3}\right)}{15 \sqrt{3} L^3},\label{rho}
\end{eqnarray}
in which the density of condensate is scaled by the bulk density, i.e., $\varrho_{\text{HIHF}}=\rho_{\text{HIHF}}/n_0$.

\begin{figure}
  \includegraphics[scale=1]{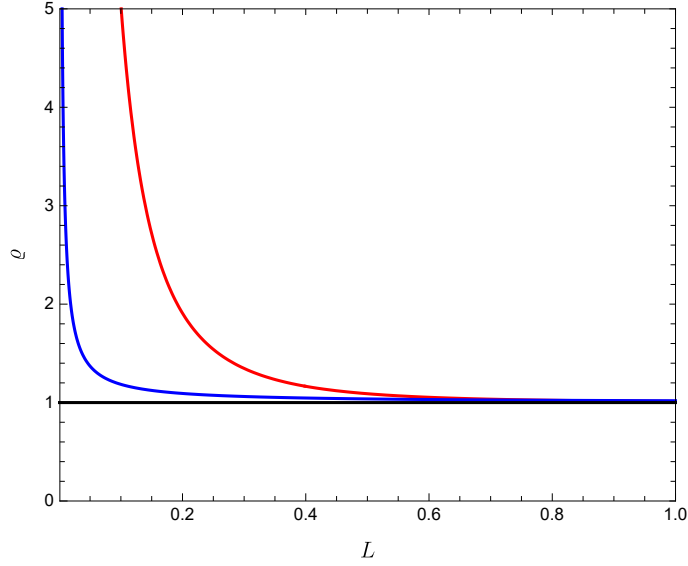}
  \caption{(Color online) The evolution of the density of state versus the distance in the HIHF approximation (red line) and LIHF approximation (blue line). The black line is the density of condensate in the one-loop approximation.}\label{f1}
\end{figure}

To illustrate for above calculations, the numerical computations are made for rubidium Rb87 $m=86.909u$, $a_s=100.4a_0$ and $\xi=4000$ {\AA} \cite{Egorov}, with $u$ and $a_0$ being atomic mass unit and Bohr radius, respectively. Fig. \ref{f1} shows the density of condensate as a function of the distance between two plates at $n_s=6.22\times 10^{-6}$. The first thing one can sees is that the density of condensate sharply decays as the distance increases, which fast approaches to unity, i.e., the bulk density $n_0$. Recall that in the LIHF approximation, $P_{11}=0$ and the second term is absent in right hand side of $P_{22}$ \cite{ThuIJMPB}. In this case it is easy to find
\begin{eqnarray}
\varrho_{{\text {LIHF}}}=1+\frac{4 \pi ^{3/2} n_s^{1/2}}{3 L}.\label{roLIHF}
\end{eqnarray}
This result is graphically shown by the blue line in Fig. \ref{f1}. It is obvious that there is no doubt about the remarkable contribution of the higher-order terms in the density of condensate. The presence of these higher-order terms makes the finite-size effect become more clearly. At a given value of the distance, the value of the density of condensate in the HIHF approximation is bigger than the one in the LIHF approximation.

\begin{figure}
  \includegraphics[scale=1]{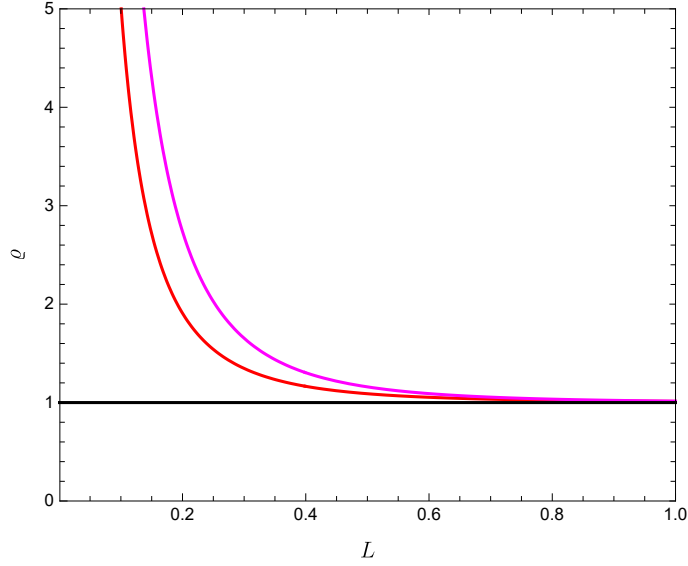}
  \caption{(Color online) The evolution of the density of state versus the distance in the HIHF approximation (red line) and square of the order parameter (magenta line). The black line is the density of condensate in the one-loop approximation.}\label{f2}
\end{figure}

A comparison between the density of the condensate and the square of the order parameter is sketched in Fig. \ref{f2}. At any value of the distance, the density of the condensate is always smaller than the square of the order parameter. This is an interesting thing, it suggests we compare the density of condensate in the CJT effective action approach with the profile of condensate in GP theory. In GP theory, the quantum fluctuations are ignored therefore one can consider the order parameter as a function of coordinate. The ground state of the Bose gas is described by the wave function, which is the solution of the GP equation and the profile (square of the wave function) is the density of condensate. In other way, in the CJT effective action approach, the coordinate dependence of the order parameter is neglected and thus it allows us to investigate the contribution of the high-order diagrams, such as, in this work all of two-loop diagrams are taken into account. This fact leads to a conclusion that at a given value of the distance, the value of the order parameter in the CJT theory equals to its value in bulk when it is calculated in the GP theory. As a consequence, the difference amount $\varrho_{\text{HIHF}}-\phi^2$ is caused by the quantum fluctuations, which corresponds to the contribution of the second term $(P_{11}+P_{22})/2$ in right hand side of Eq. (\ref{roIHF}). Fig. \ref{f2} shows that, although our system is considered at zero temperature, the quantum fluctuations are significant and its contribution in the density of condensate is not to be ignored, especially in region of the small distance. These quantum fluctuations reduce the density of condensate because of changing from the ground state to the excited state of some atoms.

To investigate the influence of the quantum fluctuations on the density of condensate one recalls the results associated with the GP theory in Ref. \cite{Biswas3}, where the planar walls locate at $z=0$ and $z=\ell$. The ground state of the Bose gas is described by the wave function $\Psi(z)$ satisfying GP equation
\begin{eqnarray}
\left[-\frac{\hbar^2}{2m}\frac{d^2}{dz^2}-\mu+gN_0\Psi(z)^2\right]\Psi(z)=0,\label{GP}
\end{eqnarray}
in which $N_0$ is the particle number. By substituting $b=2mgN_0/\hbar^2$, the GP equation (\ref{GP}) can be written as
\begin{eqnarray}
-\frac{d^2\Psi(z)}{dz^2}-\frac{1}{\xi^2}\Psi(z)+b\Psi(z)^3=0.\label{GP1}
\end{eqnarray}
Introducing the dimensionless parameters $\eta=\frac{b\xi^2}{2}n_0,~Q=\frac{\eta}{1-\eta}$ and applying the Dirichlet boundary condition at the planar walls $\Psi(0)=\Psi(\ell)=0$, the solution for (\ref{GP1}) has the form
\begin{eqnarray}
\Psi(z)=\frac{\sqrt{2Q}}{\ell\sqrt{b}}2{\mbox{EllipticK}}[Q].{\mbox{JacobiSN}}\left[2{\mbox{EllipticK}}[Q]\frac{z}{\ell},Q\right].\label{GP2}
\end{eqnarray}
It is easy to prove that the wave function $\Psi(z)$ has a maximum at the midpoint $z=\ell/2$. This fact leads to the condition
\begin{eqnarray}
\frac{\ell}{2}=\xi\frac{{\mbox{EllipticK}}\left[\frac{\eta}{1-\eta}\right]}{\sqrt{1-\eta}}.\label{GP3}
\end{eqnarray}
\begin{figure}
  \includegraphics[scale=1]{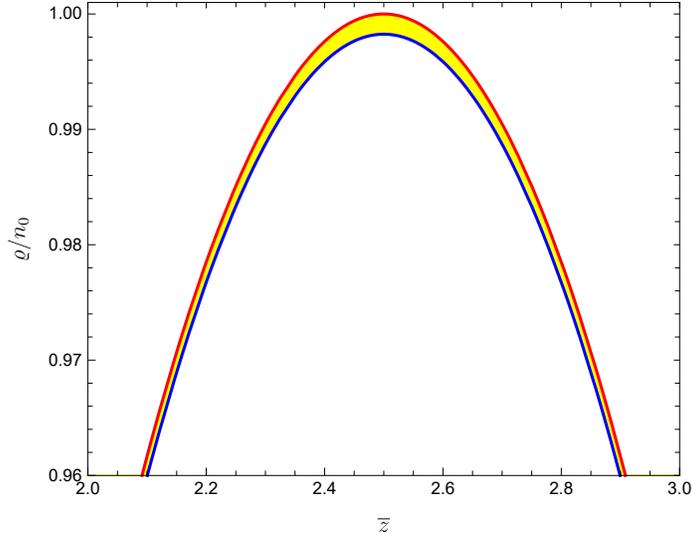}
  \caption{(Color online) Behavior of the density of condensate in the vicinity of the midpoint versus the dimensionless coordinate $\overline{z}$ at $L=5$. }\label{f3}
\end{figure}
Using dimensionless coordinate $\overline{z}=z/\xi$ and the reduced wave function $\phi_0=\Psi/\sqrt{n_0}$, the solution (\ref{GP2}) can be rewritten in from
\begin{eqnarray}
\phi_0(\overline{z})=\frac{1}{L\sqrt{1-\eta}} {\mbox{EllipticK}}[Q].{\mbox{JacobiSN}}\left[2{\mbox{EllipticK}}[Q]\frac{\overline{z}}{L},Q\right].\label{GP4}
\end{eqnarray}
In the GP theory, the density of condensate is defined as square of the wave function $\varrho_{GP}=\phi_0^2(\overline{z})$. It is shown by the red curve in Fig. \ref{f3} with the same parameter as in Fig. \ref{f1} and $L=5$. Taking into account the quantum fluctuation, the condensate density is reduced amount $\frac{(P_{11}+P_{22})}{2n_0}$ and graphically plotted by the blue curve in Fig. \ref{f3}.

\begin{figure}
  \includegraphics[scale=1]{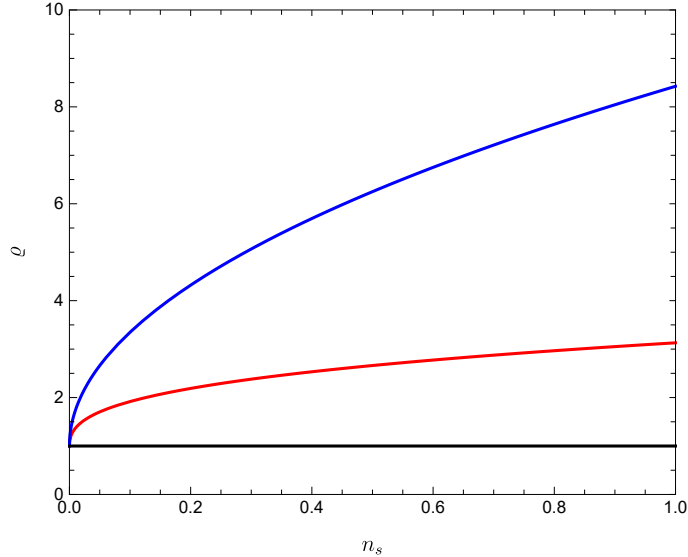}
  \caption{(Color online) The density of state as a function of the gas parameter in the HIHF approximation (red line) and LIHF approximation (blue line). The black line is the density of condensate in the one-loop approximation.}\label{f4}
\end{figure}

To end this Section, we consider the gas parameter dependence of the density of condensate. At $L=1$ this behavior is plotted in Fig. \ref{f4} by red line corresponding to the HIHF approximation and the blue line associated with the LIHF approximation. It is obvious that the density of condensate in the HIHF approximation is always smaller than the one in the LIHF approximation. This means that the higher-order terms in the momentum integrals belittle the density of condensate. It is not difficult to see that the higher-order terms in the momentum integrals contribute significantly at any value of the gas parameter. The bigger the gas parameter is, the larger the contribution of these higher-order terms in the momentum integrals is. However, for the ideal Bose gas, i.e., $n_s=0$, the density of condensate is the same in both the HIHF and LIHF approximation and equal to its value in the one-loop approximation. This means that at zero temperature the quantum fluctuations are absent in the ideal Bose gas. This result confirms again that the Casimir effect does not occur in the ideal Bose gas as pointed out in Ref. \cite{ThuIJMPB,ThuPhysa,Thu1,Biswas}.

\section{Conclusions\label{sec4}}

In the foregoing Section, the density of condensate of the dilute Bose gas confined between two planar walls is investigated in the improved Hartree-Fock approximation. The ultraviolet divergence in integrating over the perpendicular component of the wave vector is eliminated by introducing the momentum cut-off whereas the sum over all value of quantized component of the wave vector is dealt by using Euler-Maclaurin formula. Keeping up to the third order of the Euler-Maclaurin formula one arrives at the higher-order improved Hartree-Fock, meanwhile the lower-order improved Hartree-Fock corresponds to the first order of Euler-Maclaurin formula.

Our main results are in order:

- The compactification of space strongly affects on the density of condensate of the dilute Bose gas, especially in region of the small distance. The density of condensate diverges as the distance between two planar walls approaches zero.

- The contribution of these higher-order terms of the momentum integrals in the density of condensate is remarkable. This contribution is not negligible when the distance is sufficiently small.

- A comparison of the density of condensate between the CJT effective action approach and the GP theory is made. It is realized that the quantum fluctuations belittle the density of condensate in the GP theory. The resulting density of condensate in the CJT effective action approach corresponds to the one in bulk when the GP theory is employed.

Finally, our above result and the others are pointed out that the Casimir effect does not happen in the ideal Bose gas at zero temperature. This fact leads to the conclusion that the quantum fluctuations originate from the interactions among atoms in the system.

\section*{Acknowledgements}

This work is supported by Vietnam National Foundation for Science and Technology Development (NAFOSTED) and Ministry of Education and Training of Vietnam.

\section*{References}

\bibliography{mybibfile}

\end{document}